\documentclass[acus]{JAC2000}

\usepackage{graphicx}

\begin{document}
\title{\flushright{FRAT004}\\[15pt] \centering TIMING SYSTEM OF THE SWISS LIGHT SOURCE }

\author{T.Korhonen, M. Heiniger, Paul Scherrer Institute, Villigen, Switzerland}

\maketitle

\begin{abstract}
The timing system of the Swiss Light Source provides synchronization
for the accelerator and for the beamline experiments. The system is
based on an event distribution system that broadcasts the timing
information globally to all the components.
The system is based on an earlier design \cite{APS-timing} that came with an
extensive software support. We took the functionality from that
design and implemented it on a new hardware platform.
This paper describes the technical solution, the functionality of the
system and some applications that are based on the event system.

\end{abstract}

\section{INTRODUCTION}

The minimum task of a timing system for a light source is to provide 
and distribute the reference pulses to control the injection and 
accelaration.  However, from the beginning of design it was clear that we wanted
to have a system that would be more extensive.

First, good integration of the timing system into the control system
allows to bind actions to timing signals and thus makes 
it possible to implement synchronous actions in a distributed system.

Secondly, the ability to provide timestamps to collected data and 
performed actions. This makes the system to a global timebase rather 
than just a pulse delivery network.

The possibility of extending the system beyond machine timing 
was one of the major objectives for us. For example, the requirement from 
early on was to provide the means for top-up operation. This implies that 
the timing has also to be available at the beamlines.

After surveying possible solutions, the APS event system seemed to provide 
the functionality that we were looking for. The only drawback was that 
the bandwidth was not sufficient for fast timing like injection control, 
but would have required additional hardware to implement
the fast signals. However, the technology that would allow us to integrate 
all this into one subsystem was available. This prompted us to do a 
redesign of the APS system.

\section { SYSTEM OVERVIEW }

\begin{figure*}[t]
\includegraphics*[width=150mm]{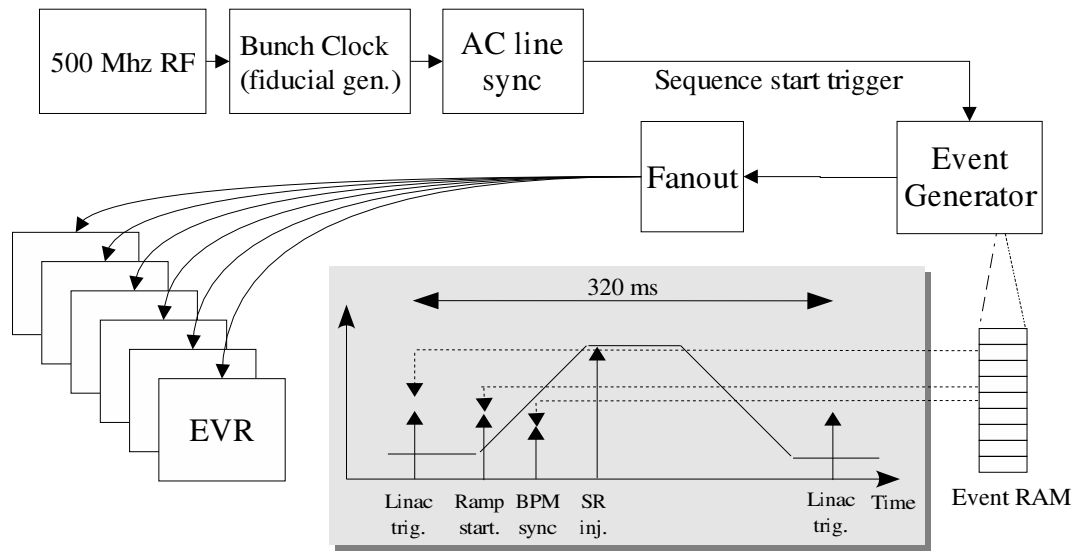}
\caption{Structure of the timing system.}
 \label{structure}
\end{figure*}

The injection system (linac and booster) of the SLS operates 
with  a 320 ms (3.125 Hz) cycle. During one cycle the linac 
is triggered, the beam is injected to the booster, the 
magnets (and RF) are ramped to accelerate the beam 
and the beam is extracted from the booster and 
injected into the storage ring after which the magnets 
are ramped down before the start of the next cycle. 
The timing system (Figure ~\ref{structure}) generates the synchronization 
reference signals and distributes these to the required components.

The reference generation is done with special purpose-built modules
like a downconverter that divides the RF frequency to the fiducial 
signals of the machine. 
The event system is used to generate the sequence of events for the injection 
cycle and to distribute the timing to the relevant components.
Practically the whole operation cycle is generated within the event generator 
from only one fiducial, namely one that tells when the number ``0'' buckets in the 
storage ring and booster are in alignment. The SLS injection sequence consists 
of about 20 events that are transmitted every 320 ms.

The event system is based on time-multiplexed transmission of event codes.
The source is an event generator that transmits 8-bit event codes at the 
a frequency of 50 MHz over an optical link. Thus, the time resolution is 20 ns.
The bitstream is multiplied to several branches with fanout modules. At 
each destination VME crate there is an event receiver that decodes the 
received event codes and performs the appropriate actions that are programmed 
for each event of interest to the particular crate.

Some subsystems like the linac have their internal timing. 
During construction the linac was required to have the 
ability to run standalone. The interface between the linac
and the rest of the machine was defined to be the timing and 
RF signals. The synchronising timing to the linac is delivered with
the event system; the linac cycle is triggered from an event.

The linac timing consists of the gun trigger system and 
the timing for the klystron modulators, prebuncher, 
gun RF amplifier and the 3 GHz frequency multiplier. 
This timing is achieved by using a pair of Stanford 
Research DG535 delay generators. The gun timing 
uses a number of TD4V delay cards \cite{TD4V-ref}, which also phase 
lock the trigger signals to the RF signal. The TD4V delay is programmable 
in units of RF cycles (2 ns) and is used to target specific 
RF buckets in the booster (and storage ring) and also 
to adjust the trigger pulse length. The trigger signal 
from these cards to the electron gun high voltage deck 
is delivered through an optical fiber using an Uniphase 
51TA/RA transmitter/receiver pair. 

\section{TECHNICAL SOLUTION}

To integrate all the desired functionality into one system, 
we decided to take the APS system as a model, but to upgrade 
to a recent technology: the Gigabit Ethernet. The requirement 
was to preserve the software compatibility so that we could take 
advantage of the software that was existing for the APS system.

The functional design was done in VHDL. This brought a big 
advantage: we could separate the hardware solution from the software 
interface, in a way that allows us a smooth upgrade path for both 
of the hardware and the software: when the communication technology 
has made an significant advance, we can port our model to the new 
technology without being tied to any particular technology or a 
component. And vice versa: any new ideas from the control system
side can be implemented in firmware without any physical modifications
(within the limits of the capabilities of the hardware.)

\subsection{The hardware}

The event system is based on Gigabit Ethernet technology, using a
standard VCSEL short wavelength (860 nm) transceiver (drop-in replacement for
long wavelength transceivers exist) and a gigabit transceiver chip. 
The transceivers allow for cascading several cards in a daisy chain.
This way we can reduce the number of fanout branches and also build a 
sub-branch where another event generator is added to an upstream branch.

The logic resides in two PLDs, one CPLD and one FPGA (Xilinx Virtex). The
CPLD handles the VME bus logic and configuration of the second one
from Flash ROM. Flash ROM allows us to upgrade the systems in situ,
even without powering down the system. This was a big advantage during 
development.

The event generator and receiver use the same PCB, with some different 
components fitted in depending on the card type. The event generator has, 
in addition to the common hardware, two 512 KB RAMs for 
``store and playback'' of an event sequence, external clock and external 
trigger inputs.
The event receiver has three extra outputs in the front panel. Their use
is defined in firmware; presently they are used for reference frequency 
output.

The main clock for the event generator is downconverted from the main 
(500 MHz) RF. The downstream receivers synchronize automatically to the 
incoming bitstream. All the components are thus phase locked to the RF.

\subsection{The firmware}

The firmware for the cards is about 2000 lines of VHDL code for both
of the modules (EVG, EVR). Being VHDL, it is independent of the
particular type of chip (FPGA) used in the hardware implementation.

The main features were already present in the APS version, we 
extended them to provide more functionality to suit our operation 
scheme.

The most essential features of the event generator firmware are 
event RAM sequence handling, event priority resolver and the possibility to send event from software by writing into a register.
The EVR has 4 channels with width and delay possibility, 14
channels with width adjustment only. 
Both of these have  also a clock prescaler to scale down the clock 
and polarity selection.
The EVR also has timestamp counter with the facilities for a 
synchronous reset and event FIFO mechanism for latching the timestamps 
when an event is received.
For details, see \cite {evt-manual}.

The firmware for the event receiver uses about 95 \% of the available
gates on the Xilinx Virtex 150.

\subsection{Interface to devices}

Different devices require different signal types for the trigger pulses. Most
devices have a TTL level input, some may have an NIM or ECL level input.
Rather than trying to cover every possible requirement on-board, we decided to put the interface logic to an transition module according to the same 
philosophy as for other SLS controls interfaces.
As an example, the power supplies for SLS \cite{andreas-psc} have an 
optical trigger input to run an internal waveform. For the power supplies, 
a transition module with optical outputs was designed.

\subsection{Distribution system}

The signals from the main timing source are distributed using an
optical fiber fanout tree. The fanout is done with arrays of VCSEL 
transmitters. These are built as VME size cards that plug in the crate, 
but have no data interface. 
One card has one input and eight outputs, limited by the card size. 
To create more outputs, one can connect the cards in a tree configuration.

\section{APPLICATIONS}

\subsection{Filling control}

The timing system must be able to precisely control the filling of the storage ring. With the event system we could create a simple application
to control the filling, without the need for any additional 
hardware and a few parameters to handle.
The harmonic numbers of the SLS booster and storage ring are 450 
and 480, respectively. This means that for each successive turn in the 
booster, the storage ring lags 30 RF buckets, until after 16 booster turns
the same buckets are aligned again. This means that by shifting the extraction
delay from the booster we can select the target bucket in 30 bucket steps.
For the smaller steps we need to adjust the linac gun timing.
Thus all that is needed to control the injection are two parameters:
extraction delay and the linac gun delay. With the proper combination of these,
all buckets in the storage ring can be reached.

The injector timing sequence is generated with the event RAM. The RAM clock
is selected to be exactly one booster turn (450/500 MHz = 900 ns)
and the cycle is started from a pulse that signals that the SR and booster 
RF buckets are in alignment. Changing the position of the
extraction event in the RAM by one the extraction can be shifted by one turn,
giving us the 30 cycle steps. The smaller steps are done by adjusting the 
linac gun timing. 
The linac timing controller is synchronized with the main timing.
Knowing the next target bucket it calculate the required delay for 
the next cycle. The calculation and setting is synchronised with the 
injection with the corresponding event: the event triggers a software 
sequence to be executed on every injection cycle. 

\subsection{Top-up}

Ability to do top-up injection was one of the requirements for the SLS from 
the beginning. This also sets several requirements for the timing system. 
The timing system has to be able to do single injection cycles, has to have 
the capability of sending gate signals to the beamline experiments and 
obviously, to synchronize several controllers that otherwise are 
interconnected only through Ethernet. The network connection 
cannot be relied on to transmit the command sequences in applications 
that are related to injection in real time. Basically, top-up is just an 
extension of the normal injection control application; it has to monitor the
beam current, allow for setting of proper injection intervals and to
have a capability to send advance notice to the beamlines before an injection.

\subsection{Beamline timing}

With our structure, we can simply extend the timing to beamlines by 
extending one fiber branch to an interested beamline, making  all the 
timing signals available at the beamline.
If the beamline has some special requirements, we can add one event
generator to create a new sub-branch to create a tree-like
structure. In this way, the beamline can add its own events but still
have the whole machine timing information available.

\section{CONCLUSIONS}

The advances in technology have made it possible to
build a timing system that can handle most of the timing tasks 
from commercial off-the-shelf components that are standardized 
and in wide use.
The SLS timing system integrates most of the required timing 
functions into a single subsystem. The system is layered so 
that we could benefit from other earlier development but at 
the same time move to a one generation newer technology. 
The functionality is captured into a device independent language 
(VHDL), which allows decoupling of the functionality from the 
underlying technology, providing a smooth upgrade path for the 
next generations.
The integration of the functions into a global system enables us
to glue a distributed system into a single entity. This can greatly 
enhance the capabilities and operation of a complex system.

\end{document}